# Quorum sensing inhibitory compounds from extremophilic microorganisms isolated from a hypersaline cyanobacterial mat

Raeid M. M. Abed[1], Sergey Dobretsov[2*], Marwan Al-Fori[2], Sarath P. Gunasekera[3], Kumar Sudesh[4] and Valerie J. Paul[3]

[1] Biology Department, College of Science, Sultan Qaboos University, Al Khoud, Sultanate of Oman

[2] Marine Science and Fisheries Department, College of Agricultural and Marine Sciences, Sultan Qaboos University, Al Khoud, Sultanate of Oman

[3] Smithsonian Marine Station, Fort Pierce, FL, USA

[4] School of Biological Sciences, Universiti Sains Malaysia, 11800, Penang, Malaysia

*Corresponding author: Sergey Dobretsov

Mailing address: Marine Science and Fisheries Department, College of Agricultural and Marine Sciences, Sultan Qaboos University, Al Khoud, Sultanate of Oman

Tel.: (968) 24143657

Fax: (968) 24413418

e-mail: sergey@squ.edu.om ; sergey_dobretsov@yahoo.com

Running title: Biotechnological potential of halophilic strains

**Abstract**

In this study extremely halophilic and moderately thermophilic microorganisms from a hypersaline microbial mat were screened for their ability to produce antibacterial, antidiatom, antialgal and quorum sensing (QS) inhibitory compounds. Five bacterial strains belonging to the genera *Marinobacter* and *Halomonas* and one archaeal strain belonging to the genus *Haloterrigena* were isolated from a microbial mat. The strains were able to grow at a maximum salinity of 22-25% and a maximum temperature of 45-60°C. Hexanes, dichloromethane and butanol extracts from the strains inhibited the growth of at least one out of nine human pathogens. Only butanol extracts of supernatants of *Halomonas* sp. SK-1 inhibited growth of the microalga *Dunaliella salina*. Most extracts from isolates inhibited QS of the acyl homoserine lactone producer and reporter *Chromobacterium violaceum* CV017. Purification of QS inhibitory dichloromethane extracts of *Marinobacter* sp. SK-3 resulted in isolation of four related diketopiperazines (DKPs): *cyclo*(L-Pro-L-Phe), *cyclo*(L-Pro-L-Leu), *cyclo*(L-Pro-L-*iso*Leu) and *cyclo*(L-Pro-D-Phe). QS inhibitory properties of these DKPs were tested using *C. violaceum* CV017 and *Escherichia coli*-based QS reporters (pSB401 and pSB1075) deficient in AHL production. *Cyclo*(L-Pro-L-Phe) and *cyclo*(L-Pro-L-*iso*Leu) inhibited QS dependent production of violacein by *C. violaceum* CV017. *Cyclo*(L-Pro-L-Phe), *cyclo*(L-Pro-L-Leu), and *cyclo*(L-Pro-L-*iso*Leu) reduced QS dependent luminescence of the reporter *E. coli* pSB401 induced by 3-oxo-C6-HSL. Our study demonstrated the ability of halophilic and moderately thermophilic strains from a hypersaline microbial mat produce biotechnologically-relevant compounds that could be used as antifouling agents.

**Key words:** Quorum sensing inhibition; antimicrobial, extreme halophiles; diketopiperazines; cyanobacterial mats; biofouling

**Introduction**

In hypersaline habitats, stratified microbial agglomerations are formed as dense thick mats, because of the limited abundance and activity of grazing organisms [23, 29]. Aerobic heterotrophic bacteria in mats are concentrated in the upper oxic layer, in close association to cyanobacteria, and belong mainly to Proteobacteria and Bacteriodetes groups [2, 30]. Strains of the genera *Marinobacter*, *Halomonas*, *Pseudomonas*, *Alcanivorax*, *Roseobacter*, *Rhodobacter* have been isolated from hypersaline mats [2]. Archaea were also detected in the oxic and anoxic layers of hypersaline microbial mats, most of which belonged to the groups Crenarchaeota, Thermoplasmatales and Euryarcheota [6, 28]. Recently, it was demonstrated that mixed bacterial communities in hypersaline cyanobacterial mats have the ability to produce bioactive compounds under *in situ* conditions [3, 21].

Halophilic microorganisms have received considerable attention because of their unique properties evolved as a result of their adaptation to high salinities. In contrast to bacteria, which produce high concentrations of compatible solutes in order to maintain an osmotic balance of their cytoplasm with the hypersaline environment, halophilic archaea accumulate high concentration of salts intracellularly. This mechanism requires special adaptation of intracellular enzymes to function in the presence of high levels of salts [15]. Therefore, extensive research has been performed to isolate compounds from halophilic bacteria, for their use as stabilizers of biomolecules and as stress-protective agents, and halophilic enzymes from archaea for their potential use as biocatalysts [39]. In contrast, little has been done to explore the production of other bioactive compounds from halophilic microorganisms with antibacterial, antialgal, antidiatom and quorum sensing (QS) inhibitory properties [42, 50].

Biofouling, defined as the quick colonization of micro- and macro-organisms on natural and artificial substrata, causes severe industrial problems, and more than $5 billion are annually spent to prevent it and to deal with its consequences [63]. Biofouling

communities develop in the form of microbial biofilms on surfaces of ships, pipelines as well as desalination units. Several antifouling compounds have been isolated from biofilm-forming microorganisms [16, 43]. Bacteria in biofilms communicate with each other and coordinate their biofilm formation by a mechanism called QS [9]. This density dependent process involves the production and release in the environment of low molecular weight compounds, such as N-acyl homoserine lactones (AHLs), which can change the behavior of bacteria after a certain threshold concentration. It has been postulated that compounds that interfere with bacterial QS can be used for antimicrobial protection [17] in aquaculture and medicine [48], as well as antifouling agents [19, 22]. Screening for antifouling compounds from halophilic microorganisms is of major industrial interest because of their stability at high salinities in marine environments and desalination plants. Additionally, these compounds can circumvent the use of toxic antifouling chemicals, like copper or zinc, which often prevent the settlement of larvae and spores of algae but not microbial biofilms [13, 20].

In this study, six microbial strains were isolated from a hypersaline cyanobacterial mat, inhabiting an inland desert wadi in South Eastern Oman. The strains were identified based on the phylogeny of their 16S rRNA genes and were physiologically and biochemically characterized. All strains were screened for the production of antibacterial, antidiatom, antialgal compounds and QS inhibitors. QS inhibitory compounds were isolated and their structures were elucidated from a representative, fast growing bacterial strain SK-3, which demonstrated high QS inhibitory activity.

**Material and Methods**

*Sample origin and isolation*

The mat samples were collected from wadi Muqshin in southeastern Oman, 200 km from the Arabian Sea coast (N 19º 35.04'; E 54º 52.87'). The wadi is a series of hypersaline, stagnant pools harbored by dome-shaped laminated microbial mats [3, 31]. Mat pieces were collected from the sides and the centre of one pool, where salinity was around 35% and air temperature was 35°C.

Small mat pieces were grown in modified growth medium (MGM) at 18% and 25% salinity [49]. The medium contained 0.1% yeast extract, 0.5% peptone, and 18% or 25% NaCl. The samples were incubated outside (temperature 25-41° C) in the shade to mimic field environmental conditions. Because of extremely high salinity of the media only 6 axenic strains (termed SK-1 to SK-6) were obtained after several transfers on MGM agar plates.

### *Biochemical characterization of the strains*

The strains were biochemically characterized for catalase, citrate utilization, indole production, methyl red test, Voges-Proskauer test, gelatin liquification, urea test and acid and gas production upon growth on the sugars D-glucose, sucrose, lactose and maltose [38]. The catalase test was performed as described by [38]. Briefly, a drop of $H_2O_2$ reagent was added on the agar plate with a few colonies of each strain; the immediate formation of bubbles indicated positive catalase activity. The strains were tested for the production of the enzymes amylase, protease, lipase, cellulase and xylanase using standard plate screening methods [11, 24]. These methods rely on the formation of clear zone in the used agar as a positive indication for the production of the enzyme. All the test plates and tubes were incubated at 25 $^{o}C$ for 3 days. Each strain was tested in triplicate.

The growth of the strains was monitored either at different salinities (0, 5, 10, 15, 18, 22, and 25%) under the constant temperature of 35 $^{o}C$ or at different temperatures (4, 15, 25, 35, 45, 60 and 70 $^{o}C$) under the constant salinity of 18%. The growth medium contained 20 mM acetate as a carbon source [2]. Growth was carried out in 10 ml glass tubes in triplicate and it was measured by following the changes of the optical densities of the cultures at 660 nm. Bacterial culture in the medium without acetate and the sterile medium with acetate but without bacteria were used as controls.

*Strain identification*

The DNA of the isolates was extracted and the 16S rRNA genes were PCR amplified and then sequenced as previously described [2]. The 16S rRNA sequences of the strains (>1300 bp) were analyzed using the ARB software [47]. Phylogenetic trees were constructed based on almost complete 16S rRNA gene sequences (>1300 bp) by applying different methods integrated in the ARB software such as maximum likelihood, maximum parsimony and neighbour-joining. Partial sequences were not included in the calculation of the trees but were inserted afterwards using the parsimony tool. The final maximum likelihood tree was minimized for simplicity in presentation.

*Preparation of extracts*

Preliminary results suggested that the cell cultures grown for 7 days on MGM agar had lower bioactivity and 3-fold lower yield of extracts in comparison with ones grown in MGM liquid medium. Therefore only results for MGM liquid medium are presented. The cells cultures grown for 7 days in MGM liquid medium (18% NaCl, volume 250 ml) at 37 $^{o}$C were separated by centrifugation (5000 *g* for 10 minutes) and the pellets were re-suspended in 100 ml ddH$_2$O. Supernatants and pellet suspensions were subsequently partitioned over 3 days using a separatory funnel. Hexanes (a mixture of isomers of C6 containing aliphatics including hexane, cyclohexane and branched pentanes), dichloromethane and butanol (all from Sigma-Aldrich) were used as solvents for partitioning. Resulting hexanes, dichloromethane, butanol and remaining aqueous fractions were individually collected, filtered through Whatman paper No.1 and evaporated by a rotary evaporator (Büchi) under reduced pressure at 35 $^{o}$C. All dry extracts were weighed, stored in glass vials at -20 $^{o}$C and used later for bioassays (see below).

*Antibacterial, antidiatom and antialgal activities*

Antibacterial activity of the obtained extracts was tested against the pathogenic bacteria *Bacillus subtilis*, *Escherichia coli*, *Micrococcus luteus*, *Proteus vulgaris*, *Pseudomonas aeruginosa*, *Salmonella enterica*, *Shigella sonnei*, *Staphylococcus aureus* and *Streptococcus pyogenes*. All strains were obtained from the culture collection of the Sultan Qaboos University hospital. *E.coli*, *P. vulgaris* were tetracycline and streptomycin resistant. *P. aeruginosa* strain was tetracycline resistant, while the *S sonnei* strain was streptomycin resistant. Prior to the bioassay, all strains were cultivated for 15h in Luria-Bertani (LB) broth (Fisher Scientific) at 37 $^{o}$C . All extracts were re-dissolved in appropriate solvents in order to reach the final concentration of 1 mg ml$^{-1}$. Antibacterial bioassays were conducted according to [18]. Briefly, each strain (0.1 ml of 10$^6$ CFU ml$^{-1}$ culture) was inoculated onto LB agar plates. Three µl of the extracts or control solvents were applied to sterile paper disks (diameter 0.5 cm) made of Whatman No.1 filter paper. The discs were dried (resulting in the final amount of 3 µg extract per disk) and placed onto the agar plates. The agar plates were incubated at 30° C for 24h and the formation of inhibition zones in triplicate was measured to the nearest 0.2 mm. The mean inhibition zones in the presence of extracts were compared with the control (no extract) using the Dunnet test (ANOVA) [65]. To achieve normality prior to the analysis, data were square root transformed.

Antidiatom and antialgal activities were tested using the diatom *Amphora coffeaeformis* (*Bacillariophyceae*) isolated from a natural biofilm [44] and the green alga *Dunaliella salina* CCAP 19/18, respectively, according to [21] with some modifications. Prior to the experiment the diatoms and alga were cultivated in aerated Erlenmeyer flasks filled with 400 ml of F/2 medium at 25$^o$ C with continuous overhead illumination at 0.3 × 10$^{16}$ quanta s$^{-1}$ cm$^{-1}$. Three µl of the extracts (concentration 1 mg ml$^{-1}$) or control solvents were applied to sterile paper disks (diameter 0.5 cm) made of Whatman No.1 filter paper. The discs were dried and each disk was placed in a separate well of 24-well plates (Corning).

Using a sterile brush, a suspension of algae or diatoms was made and 1 ml of this suspension was individually applied into each well of the multi-well plate. After 3 days of incubation with continuous illumination ($0.3 \times 10^{16}$ quanta s$^{-1}$ cm$^{-1}$) at 25°C, the amount of chlorophyll *a* (µg L$^{-1}$) in each well was determined by the technique of [37] using a spectrophotometer (Beckman). Each experiment was run in triplicate. The mean amount of chlorophyll *a* in the presence of extracts was compared with the control (the sterile disks without extract) using the Dunnet test (ANOVA). To achieve normality prior the analysis, data were square root transformed.

*Screening of QS inhibitory compounds*

*Chromobacterium violaceum* CV017 was used as a reporter strain for screening QS inhibitory properties of the extracts. The strain produces N-hexanoyl homoserine lactone, which induces production of the purple pigment violacein via the AHL receptor CviR; it has been obtained after a spontaneous mutation of the wild-type *C. violaceum* ATCC 31532 [14]. AHLs with acyl side chains >C6 antagonize CviR in this reporter. Experiments were conducted according to [22]. Briefly, compounds at concentrations of $3 \times 10^{-6} - 3 \times 10^{-2}$ mg ml$^{-1}$ were individually applied into wells of microtiter plates (Nunc) and solvents were evaporated under the air flow in the laminar hood. Then, extracts were re-dissolved in 1 µl of dimethyl sulfoxide (DMSO). DMSO in empty cells was used as a control. Five ml of soft LB agar (Difco) were mixed with 500 µl of washed overnight culture of CV017, and 100 µl of this mixture (90.9 of LB agar and 9.1 µl of CV017) were applied to each well. The plates were incubated 24h at 30 °C. Due to the colour of different extracts, which can interfere with the plate-based violacein measurement method, a reduction in violacein production was compared to the control treatments visually [22]. The bioassays were repeated three times and the mean minimum inhibitory concentration that inhibited QS (MIC) in µg ml$^{-1}$ was calculated. Possible toxic effects of extracts on the CV017 reporter were investigated according to [22]. The MICs in the presence of extracts were compared with the control (DMSO) using the Dunnet test (ANOVA).

*Isolation and identification of QS inhibitory compounds from SK-3*

Strain SK-3 was selected for isolation of QS inhibitory compounds due to its high growth rate and extract yield as well as QS bioactivity. The strain was grown for 7 days in 5 L of MGM liquid medium with 15% NaCl at 37 $^{o}$C. The supernatants were separated from cell pellets using centrifugation (5000 *g* for 10 minutes) and the supernatants were further extracted with dichloromethane for 3 days. The dichloromethane fraction was evaporated using a rotary evaporator under reduced pressure at 37 $^{o}$C. Dichloromethane extract (weight = 1g) was purified by a silica column (weight of silica =15g) and the column was eluted sequentially with 250 ml of 70:30% hexanes: ethyl acetate, 100% ethyl acetate, 10:90% methanol: ethyl acetate, and 100% methanol. All solvents were HPLC grade obtained from Fisher Scientific. Fractions were collected separately and evaporated using a rotary evaporator. Each collected fraction was re-dissolved either in 1 ml of ethyl acetate (fractions 1, 2 and 3) or methanol (fraction 4). These fractions were tested for their ability to inhibit QS of *C. violaceum* CV017. The active fraction 3 (10:90% methanol: ethyl acetate; dry weight = 74 mg) was dissolved in 1 ml of 9:1 MeOH: water and separated using a C-18 pre-packed mini Sep-pac column (Waters). This column was eluted with 90% methanol and 100% methanol to yield 5 fractions. Each fraction was collected separately, evaporated and assayed using *C.violaceum* CV017. The active fraction (90% methanol, weight 10 mg) was further purified using a preparative reverse phase HPLC (Waters, C18 YMC-pack column, 5μ, length =250 mm). A twenty minute HPLC run with 43% methanol/57% water and a flow rate = 3 ml min$^{-1}$ resulted in isolation of 1.5 mg of diketopiperazine (**1**) *cyclo*(L-Pro-L-Phe) (Fig. 3). Remaining material was separated using a preparative reversed phase HPLC (Waters, C18 YMC-pack column, 5μ, length =250 mm), run with 60% methanol and a flow rate = 3 ml min$^{-1}$ for 20 minutes. This resulted in isolation of 1 mg of diketopiperazine (**2**) *cyclo*(L-Pro-L-Leu), 0.2 mg of diketopiperazine (**3**) *cyclo*(L-Pro-L-*iso*Leu) and 0.6 mg of diketopiperazine (**4**) *cyclo*(L-Pro-D-Phe) (Fig.3). Compounds were identified by comparison with existing $^{13}$C and $^{1}$H NMR, specific rotation data and LREIMS data.

*Screening of QS inhibitory compounds from SK-3*

Before the bioassays, all diketopiperazines (DKPs) from SK-3 were re-dissolved in DMSO. As a control DMSO was used. In order to investigate in detail QS inhibitory properties of isolated compounds we used three reporters. An AHL producer and reporter strain *C. violaceum* CV017 changes its colour in response to short <C5 acyl side chains. Two other *E.coli*-based QS reporters were deficient in AHL production; reporter *E. coli* pSB401 [61] that contained the *luxR* P$_{luxI}$-*luxCDABE* transcriptional fusion and emits light in response to AHLs with medium C6-C8 acyl side chains, and LasR-based *E.coli* reporter pSB1075 [61] that contained the *lasR* P$_{lasI}$-*luxCDABE* and emits light in response to AHLs with long >C10 acyl side chains. Possible toxic effects of compounds on metabolic activity or luminescence of the reporters were tested using a control construct containing a pTIM2442 plasmid in *E. coli* DH5α [4]. pTIM2442 carries the *luxCDABE* cassette controlled by a constitutive phage lambda promoter. In a case of toxicity of tested compounds light production by this reporter is reduced. Direct and indirect experiments were performed with 4 replicates according to [22]. For all bioassays, DKPs were added to the wells of a black microtiter plate (Nunc) and reporter suspensions in LB broth were added. In the direct bioassay, AHLs N-3-oxo-hexanoyl-L-homoserine lactone (3-oxo-C6-HSL), final concentration of 10 μM, or N-3-oxo-dodecanoyl-L-homoserine lactone (3-oxo-C12-HSL), final concentration of 2 μM, were added to the wells of a black microtiter plate in order to stimulate QS of pSB401 or pSB1075, correspondingly. In indirect bioassays, different concentrations of compounds were exposed to the reporters without AHLs. Luminescence and optical density of the reporter suspensions (OD$_{595}$) were measured every hour using a multimode microtiter plate reader Victor-3 (Perkin Elmer). The bioluminescence data are presented as "relative bioluminescence" (RB) in order to take into account the population density of the reporters according to the following formula:

RB= (B$_s$)/ OD$_{595}$

Where B$_s$ is bioluminescence of each sample measured in counts per second (CPS) using the plate reader, and OD$_{595}$ is optical density of the reporter culture measured at 595 nm. The differences between the treatments and the positive control were compared by ANOVA followed by a Dunnet test [65].

**Results**

***Morphological, phylogenetic and physiological characteristics of the strains***

All strains have rod-shaped cells, 0.5-0.75 μm in diameter and 2-5 μm length, with terminal flagellae except for the strain SK-6. 16S rRNA-based phylogeny placed 5 strains (SK-1-5) within the *Gammaproteobacteria* while SK-6 belonged to Archaea (Fig. 1). The strains SK-2, SK-4 and SK-5 were phylogenetically related to sequences of the genus *Halomonas*, while SK-1 and SK-3 fell within sequences of the genus *Marinobacter*. The closest relative to the strains SK-1 and SK-3 was *Marinobacter haloterrigenus* with more than 99% sequence similarity. The strain SK-6 shared around 99.6% sequence similarity with the archaeons *Haloterrigena saccharevitans* isolated from Aibi salt lake, Xin-Jiang, China [62] and *Haloterrigena thermotolerans* isolated from the solar salterns of Cabo Rojo, Puerto Rico [40].

All strains were Gram negative and were negative for methyl red (MR), Voges–Proskauer (VP), urea, citrate, gelatin liquefication and triple sugar iron (TSI) tests (Table 1).

Only SK-1 and SK-3 showed catalase activity. Acid production from D-glucose, lactose, and maltose was only observed in the case of SK-4 and SK-6, whereas SK-5 showed acid production only on D-glucose and SK-2 only on lactose. All strains could hydrolyze at least two of the compounds; tween 20 and carboxymethyl cellulose (CMC) or tween 20 and Birchwood xylan, except SK-6 which could not hydrolyse any of them (Table 1). All strains grew well at temperatures between 15-45°C and not below 5°C or above 50°C, except SK-6, which could not grow below 20°C or above 60°C (Table 1). The lowest optimal temperature was observed for the strain SK-5 (25 °C), while the highest one was found for the strain SK-6 (45 °C). All strains grew at salinities between 5-22%, but none grew at 0%. SK-6 could not grow at salinities below 15% and tolerated salinities up to 25% (Table 1).

None of the strains produced amylase and protease, however, all strains except SK-6 produced lipase (Table 1). While cellulase production was only observed in the case of the strains SK-1, SK-2 and SK-3, xylanase was only produced by SK-2, SK-4 and SK-5.

*Antibacterial activity*

The growth of four out of nine tested pathogens (*Bacillus subtilis*, *Staphylococcus aureus*, *Streptococcus pyogenes* and *Salmonella enterica*) was significantly inhibited (ANOVA, Dunnet test, $p<0.05$) by extracts from SK-3 (Table 2). Extracts of the other isolates had an inhibitory effect on three tested pathogens only. All hexanes extracts of the isolates inhibited growth of *B. subtilis*; however, hexanes extracts of supernatants were more active than cell pellet extracts (Table 2). All dichloromethane extracts, except the supernatant of SK-1, inhibited growth of the pathogen *S. aureus*. A relatively large inhibition zone of 6.0 ±1.1 mm in diameter was obtained in the case of SK-2 dichloromethane pellet extract. Butanol extracts of all cell pellets, but not their supernatants, inhibited the growth of *S. pyrogenes*. Only the butanol pellet extract of SK-3 inhibited *S. enterica* and *B. subtilis*, additionally. Water extracts were not active against any of the tested pathogens.

*Antidiatom and antialgal activity*

Screening for antidiatom bioactivity showed variable effects among the different extracts and different strains (Fig.2). While dichloromethane and butanol extracts of supernatants of SK-1 significantly (ANOVA, Dunnet test, p<0.05) inhibited growth of the diatom *Amphora coffeaeformis*, hexanes extracts from cell pellets of SK-2, SK-3 and SK-6 and supernatants and water extracts from cell pellets of SK-4 significantly enhanced its growth (Fig. 2). In the remaining cases, there was no detectable effect. In the case of the green alga *Dunaliella salina*, only butanol extracts of supernatants of SK-1 significantly (ANOVA, Dunnet, P<0.05) inhibited its growth (Fig. 2A). In contrast, supernatants and hexanes extracts of supernatants of SK-2, SK-4 and SK-6, hexanes extracts of cell pellets from SK-2, SK-3 and SK-6 and water extracts from pellets of SK-5 significantly (ANOVA, Dunnet, P<0.05) induced growth of *D. salina*.

*QS inhibition by extracts of isolates*

All supernatants and water extracts of cell pellets of all strains significantly (ANOVA, Dunnet test, p<0.05) inhibited QS of the *C. violaceum* CV017 strain (Table 3). Supernatant of SK-5 had the highest bioactivity (i.e. the lowest MIC). All extracts of SK-1 and SK-3 inhibited QS of the reporter (Table 3). QS inhibitory activity was higher in SK-3 than SK-1 extracts. There was no toxic effect of tested extracts on the reporter strain, and the DMSO control did not affect QS of the reporter strain (data are not shown).

*Inhibition of QS by compounds isolated from SK-3*

Four DKPs (**1**) *cyclo*(L-Pro-L-Phe), (**2**) *cyclo*(L-Pro-L-Leu), (**3**) *cyclo*(L-Pro-L-*iso*Leu) and (**4**) *cyclo*(L-Pro-D-Phe) were isolated and identified from the bacterium SK-3 (Fig.3). In experiments with *C. violaceum* CV017, *cyclo*(L-Pro-L-Phe) and *cyclo*(L-Pro-L-*iso*Leu) inhibited QS dependent production of violacein at 160 μM and 94 μM, respectively (data are not shown). These compounds were not toxic to this reporter strain at tested concentrations (data are not shown). In the case of the *E. coli* pTIM2442 reporter, only *cyclo*(L-Pro-L-Phe) significantly (ANOVA, Dunnet test, $p<0.05$) inhibited luminescence at 80 μM (data are not shown). The QS dependent luminescence of the reporter *E. coli* pSB401 induced by 3-oxo-C6-HSL was significantly (ANOVA, Dunnet test, $p<0.05$) inhibited by *cyclo*(L-Pro-L-Phe), *cyclo*(L-Pro-L-Leu), and *cyclo*(L-Pro-L-*iso*Leu) (Fig. 4). While *cyclo*(L-Pro-L-Phe) inhibited QS of the reporter *E. coli* pSB401 at 20-40 μM, its D-isomer *cyclo*(L-Pro-D-Phe) did not have any activity (Fig.4). Relative bioluminescence of the negative control (background relative bioluminescence of pSB401 without 3-oxo-C6-HSL) was very low (Fig.4). None of the isolated DKPs affected QS dependent luminescence of the reporter *E. coli* pSB1075 induced by 3oxo-C12-HSL or without it (data are not shown).

**Discussion**

Halophilic microorganisms possess a multitude of biotechnologically relevant bioactive compounds, produced as a result of their unique physiological and genetic properties evolved to cope with hypersaline conditions [39]. Due to extremely high salinity only six strains belonging to known halophiles of the genera *Marinobacter*, *Halomonas* and *Haloterrigena* were isolated. Isolates belonging to *Marinobacter* and *Halomonas* have been previously isolated from hypersaline microbial mats [2, 30, 33, 57], whereas *Haloterrigena*-related species were mainly obtained from crystallizer ponds of solar saltens [40, 62]. The ability of our bacterial and archaeal strains to grow at salinities up to 22% and 25% and temperatures up to 45 $^0$C and 60 $^0$C, respectively, classifies them as extreme halophiles and moderate thermophiles.

Extracts of our halophilic isolates inhibited growth of some pathogens (Table 2). All observed antibacterial activities were found in the non-polar (butanol, dichloromethane and hexanes) and not in the polar (water) fractions. However, it remains unknown whether this high antibacterial activity of non-polar extracts is a unique property of our strains or a general attribute of other halophilic microorganisms. Additionally, it is not clear if the same bioactivity could be observed for these strains growing in microbial mats under natural conditions. For example, cultivation of our isolates in the liquid and solid media under laboratory conditions affected their bioactivity. Although all tested strains inhibited the growth of *B. subtilis*, *S. aureus* and *S. pyrogenes*, only *Marinobacter* sp. SK-3 inhibited the growth of *S. enterica*. For the first time in this study, we reported the ability of *Haloterrigena* isolates to produce antibacterial compounds. Ethyl acetate extracts of a *Marinobacter* sp. associated with sponges was shown to inhibit the growth of *B. subtilis* [7], and a marine *Halomonas* sp. was shown to produce aminophenoxazinones with antibacterial properties [12]. Many representatives of the family *Halobacteriaceae* were demonstrated to produce bacteriocins [32]. Despite the fact that many halobacteria produce bacteriocins in cultures, these compounds could not be detected in the field. This suggests that the role of bacteriocins in the competition between different halobacteria in hypersaline aquatic environments is probably negligible [32].

To the best of our knowledge, no antialgal or antidiatom activities have been recorded from halophilic microbes, and our isolates are no exception. Only *Halomonas* sp. SK-1 inhibited the growth of the green alga *Dunaliella salina* and the diatom *Amphora coffeaeformis* and a few of the extracts induced their growth (Fig. 2). Possibly, these results could be explained by the low levels of competition between bacteria, archaea, microalgae and diatoms in hypersaline mats. In contrast, the lack of antialgal activity could be due to the selection of target strains for our bioassay. The induction of growth of *D. salina* and *A. coffeaeformis* is an interesting aspect and may point to the presence of stimulatory compounds in our extracts. Indeed, it was previously shown that the presence of aerobic heterotrophs induces the growth of phototrophic organisms, such as cyanobacteria [1]. The addition of the bacterium *Marinobacter adhaerens* HP15 was shown to enhance the aggregate formation of the diatom *Thalassiosira weissflogii* [53]. Cyanobacteria and associated heterotrophic bacteria were shown to exchange vitamins, other growth factors, nitrogen and carbon sources, leading to enhanced cyanobacterial growth [54]. The filtrate of aerobic bacterial cultures lacking microbes was shown to have a stimulatory effect on the growth of cyanobacteria (K. Khols, personal communication), indicating that these bacteria produced chemicals that account for this effect. More research is still needed to understand the complex interaction between different organisms in microbial mats and the nature of chemicals produced that regulate these relationships under natural conditions.

In this study, halophilic bacterial and archaeal isolates from a microbial mat produced compounds with the ability to inhibit QS dependent production of violacein by *C. violaceum* CV017 (Table 3). Ethyl acetate extracts of total hypersaline mats were previously shown to exhibit some QS inhibitory properties against *Agrobacterium tumefaciens* NTL4 (pZLR4) and *Salmonella enterica* S235, but not *C. violaceum* CV017 [3]. Inhibition of *Pseudomonas aeruginosa* PAO1 biofilm formation and QS dependent violacein production of *C. violaceum* strain CV026 by isolates from marine sediments belonging to different genera including *Marinobacter* has been reported [41]. An isolate belonging to *Marinobacter* inhibited both biofilm formation and swarming in *Serratia marcescens* but did not affect its growth [5], suggesting production of multiple chemical inhibitors. Inhibition of QS by compounds produced by *Halomonas* spp. and archaean strains has never been reported before. In contrast, the production of QS compounds that activate bacterial N-acyl homoserine lactone (AHL) bioreporters by *Halomonas* spp. as well by the archaeal species *Haloterrigena hispanica* and *Natronococcus occultus* has been reported [36, 45, 55, 56].

We isolated several diketopiperazines (DKPs) from *Marinobacter* sp. SK-3 (Fig. 3) and demonstrated their QS inhibitory activities (Fig.4). While DKPs have been isolated from bacteria belonging to the genera *Bacillus* [59, 64], *Streptomyces* [34], *Pseudomonas* [25],

*Pseudoalteromonas* [51], *Burkholderia* [59], and *Microbispora* [27], the presence of DKPs in *Marinobacter* sp. has not been previously reported. It is not clear whether *Marinobacter* sp. SK-3 produces these DKPs under natural conditions at concentrations sufficient to inhibit QS of other bacteria and this should be investigated in future experiments. It was shown that the bacterium *Burkholderia cepacia* uses *cyclo*(Pro-Phe), *cyclo*(Pro-Tyr), *cyclo*(Ala-Val), *cyclo*(Pro-Leu) and *cyclo*(Pro-Val) as signal QS molecules [59]. In contrast, our study suggested that *cyclo*(L-Pro-L-Phe), *cyclo*(L-Pro-L-*iso*Leu) and *cyclo*(L-Pro-L-Leu) act as inhibitors of bacterial QS (Fig. 4). Similarly, other DKPs, such as *cyclo*(Delta-Ala-L-Val), *cyclo*(L-Pro-L-Tyr) and *cyclo*(L-Phe-L-Pro), competed with the natural inducer 3-oxo-C6-HSL and inhibited QS in several reporter strains [25]. DKPs were also detected in fungi [8, 35, 46, 60], as well as in the extremely halophilic archaeon *Haloterrigena hispanica* [56]. In the latter organism, the compound *cyclo*(L-Pro-L-Val) was able to induce QS in bacterial AHL reporters, suggesting the ability of archaea to interact with AHL-producing bacteria within mixed communities. This DKP is different from those isolated from our SK-3 strain. The broad distribution of DKPs among microorganisms, and their ability to activate bacterial QS and inhibit this process, points to a vital role of these compounds in microbial communities.

The effect of isolated DKPs on different reporters was different. QS-dependent production of violacein by *C. violaceum* CV017 (responses to AHL with short <C5 acyl side chains) was affected by *cyclo*(L-Pro-L-Phe) and *cyclo*(L-Pro-L-*iso*Leu), while QS-dependent luminescence of the reporter *E. coli* pSB401 (emits light in response to AHLs with medium C6-C8 acyl side chains) was inhibited by *cyclo*(L-Pro-L-Phe), *cyclo*(L-Pro-L-Leu), and *cyclo*(L-Pro-L-*iso*Leu) (Fig. 4). None of isolated DKPs affected QS dependent luminescence of the reporter *E. coli* pSB1075 (responses to AHLs with long >C10 acyl side chains). Isomeric configuration of isolated DKPs determines their bioactivity. For example, *cyclo*(L-Pro-L-Phe) had some QS inhibitory properties, while its isomer *cyclo*(L-Pro-D-Phe) did not have any activity. *Cyclo*(L-Pro-L-*iso*Leu) at a concentration of 80 μM inhibited *E. coli* pTIM2442 reporter luminescence, suggesting that at this concentration it was either toxic to the reporter or inhibited its luminescence directly (i.e. by affecting the luciferase enzyme) or indirectly (by affecting metabolism). It was demonstrated that DKPs have different bioactivities including antiviral, antimicrobial, antifungal, anticancer, cytotoxic and antifouling [10, 26, 34, 51, 52, 58]. Since DKPs have multiple biological actions, these compounds have huge potential as antifouling agents, not only because they can suppress bacterial QS and inhibit bacterial growth but can also prevent settlement of invertebrate larvae.

In conclusion, our halophilic strains from hypersaline microbial mats have the ability to produce antibacterial and QS inhibitory compounds. DKPs from *Marinobacter* sp. SK3 can control bacterial QS and may potentially be used as antifouling agents to solve problems caused by biofilm formation in marine environments and desalination plants, where salinity is often high. Microbial isolates from hypersaline mats seem to be an important source of bioactive substances, with potential applications in medicine, industry and environmental settings.

**Acknowledgements**

The work of SD was supported by a Sultan Qaboos University (SQU) internal grant IG/AGR/FISH/12/01, by a HM Sultan Qaboos Research Trust Fund SR/AGR/FISH/10/01 and by the George E. Burch Fellowship in Theoretical Medicine and Affiliated Sciences at the Smithsonian Institution (USA). SD acknowledge Professor Ricardo Coutinho (IEAPM, Arraial do Cabo, Brazil) and the program science without frontiers (CNPq). RA would like to thank the Hanse-Wissenschaftskolleg (HWK), Institute for Advanced Study, Germany for their support. The authors acknowledge help with QS bioassays by Dr. Max Teplitski (University Florida, USA) and with strain isolation by Mrs. Samiha Al Kharusi.

**References**

1. Abed RMM, Zein B, Al-Thukair A, de Beer D (2003) Phylogenetic diversity and activity of aerobic heterotrophic bacteria from a hypersaline oil-polluted microbial mat. Aquat Microb Ecol 30: 127-133

2. Abed RMM, Zein B, Al-Thukair A, de Beer D (2007) Phylogenetic diversity and function of aerobic heterotrophic bacteria from a hypersaline oil-polluted microbial mat. Syst Appl Microbiol 30: 319-330

3. Abed RMM, Dobretsov S, Al-Kharusi S, Schramm A, Jupp B, Golubic S (2011) Cyanobacterial diversity and bioactivity of inland hypersaline microbial mats from a desert stream in the Sultanate of Oman. Fottea 11: 215-224

4. Alagely A, Rajamani S, Teplitski M (2011) Luminescent reporters and their applications for the characterization of signals and signal-mimics that alter LasR-mediated quorum sensing. Methods Mol Biol 692: 113-130

5. Alagely A, Krediet CJ, Ritchie KB, Teplitski M (2011) Signaling-mediated cross-talk modulates swarming and biofilm formation in a coral pathogen *Serratia marcescens*. ISME Journal 5: 1609-1620

6. Allen MA, Goh F, Burns BP, Neilan BA (2009) Bacterial, archaeal and eukaryotic diversity of smooth and pustular microbial mat communities in the hypersaline lagoon of Shark Bay. Geobiology 7: 82-96

7. Anand TP, Bhat AW, Shouche YS, Roy U, Siddharth J, Sarma SP (2006) Antimicrobial activity of marine bacteria associated with sponges from the waters off the coast of South East India. Microbiol Res 161: 252-262

8. Antia BS, Aree T, Kasettrathat C, Wiyakrutta S, Ekpa OD, Ekpe UJ, Mahidol C, Ruchirawat S, Kittakoop P (2011) Itaconic acid derivates and diketopiperazine from the marine-derived fungus *Aspergillus aceleatus* CRI322-03. Phytochemistry 72: 816-822

9. Antunes LC, Ferreira RB, Buckner MM, Finlay BB (2010) Quorum sensing in bacterial virulence. Microbiology 156: 2271-2282

10. Bertinetti BV, Peña N, Cabrera GM (2009) An antifungal tetrapeptide from the culture of *Penicillum canescens*. Chem Biodivers 6: 1178-1184

11. Bragger JM, Daniel RM, Coolbear T, Morgan HW (1989) Very stable enzymes from extremely thermophilic archaebacteria and eubacteria. Appl Microb Biotech 31: 556-561

12. Bitzer J, Grosse T, Wang L, Lang S, Beil W, Zeeck A (2006) New aminophenoxazinones from a marine *Halomonas* sp.: fermentation, structure elucidation, and biological activity. J Antibiotics 59: 86-92

13. Cassé F, Swain GW (2006) The development of microfouling on four commercial antifouling coatings under static and dynamic immersion. Int Biodeter Biodegr 57: 179-185

14. Chernin LS, Winson MK, Thompson JM, Haran S, Bycroft BW, Chet I, Williams P, Stewart GS (1998) Chitinolytic activity in *Chromobacterium violaceum*: substrate analysis and regulation by quorum sensing. J Bacteriol 180: 4435- 4441

15. Da Costa MS, Santos H, Galinski EA (1998) An overview of the role and diversity of compatible solutes in *Bacteria* and *Archaea*. Biotech Extremophiles 61: 117- 153

16. Dash S, Jin C, Lee OO, Xu Y, Qian P-Y (2009) Antibacterial and antilarval-settlement potential and metabolite profiles of novel sponge-associated marine bacteria. J Ind Microb Biotech 36: 1047-1056

17. Defoirdt T, Boon N, Bossier P, Verstraete W (2004) Disruption of bacterial quorum sensing: An unexplored strategy to fight infections in aquaculture. Aquaculture 240: 69-88

18. Dobretsov S, Qian PY (2002) Effect of bacteria associated with the green alga *Ulva reticulata* on marine micro- and macrofouling. Biofouling 18: 217-228

19. Dobretsov S, Teplitski M, Paul JV (2009) Mini-review: quorum sensing in the marine environment and its relationship to biofouling. Biofouling 25: 413-427

20. Dobretsov S, Thomason J (2011) The development of marine biofilms on two commercial non-biocidal coatings: a comparison between silicone and fluoropolymer technologies. Biofouling 27: 869-880

21. Dobretsov S, Abed RMM, Al Maskari SMS, Al Sabahi JN, Victor R (2011) Cyanobacterial mats from hot springs produce antimicrobial compounds and quorum sensing inhibitors under natural conditions. J Appl Phycol 23: 983-993

22. Dobretsov S, Teplitski M, Bayer M, Gunasekera S, Proksch P, Paul VJ (2011). Inhibition of marine biofouling by bacterial quorum sensing inhibitors. Biofouling 27: 893-905

23. Farmer JD (1992) Grazing and bioturbation in modern microbial mats. In: Schopf JW, Klein C (eds) The Proterozoic Biosphere - a Multidisciplinary Study. Cambridge, Cambridge University Press, pp 247-251

24. Haba E, Bresco O, Ferrer C, Marques A, Busquests M, Manresa A (2000) Isolation of lipase-secreting bacteria by deploying used frying oil as selective substrate. Enzyme Microb Tech 26: 40-44

25. Holden MT, Chhabra SR, de Nys R, Stead P, Bainton Nj, Hill PJ, Manefield M, Kumar N, Labatte M, England D, et al (1999) Quorum-sensing cross talk: isolation and chemical characterization of cyclic dipeptides from *Pseudomonas aeruginosa* and other Gram-negative bacteria. Mol Microbiol 33:1254-1266

26. Huang R, Zhou X, Xu T, Yang X, Liu Y (2010). Diketopoperazines from marine organisms. Chem Biodivers 7: 2809-2829

27. Ivanova V, Laatsch H, Kolarova M, Aleksieva K (2013) Structure elucidation of a new natural diketopiperazine from a *Microbispora aerate* strain isolated from Livingston Island, Antarctica. Nat Prod Res 27: 164-170

28. Jahnke LL, Orphan VJ, Embaye T, Turk KA, Kubo MD, Summons RE, Des Marais DJ (2008) Lipid biomarker and phylogenetic analyses to reveal archaeal biodiversity and distribution in hypersaline microbial mat and underlying sediment. Geobiology 6: 391-410

29. Javor BJ, Castenholz RW (1984) Invertebrate grazers of a microbial mat, Laguna Gurrero Negro, Mexico. In: Cohen Y, Castenholz RW, Halvorson HO (eds) Microbial Mats: Stromatolites. New York, Alan R Liss Inc, pp 85-94

30. Jonkers HM, Abed RMM (2003) Identification of aerobic heterotrophic bacteria from the photic zone of a hypersaline microbial mat. Aquat Microb Ecol 30: 127-133

31. Jupp BP, Eichenberger U, Cookson P (2008) The microbial domes of Wadi Muqshin pools, Sultanate of Oman. Int J Environ Stud 65: 685-703

32. Kis-Papo T, Oren A (2000) Halocins: are they involved in the competition between halobacteria in saltern ponds? Extremophiles 4: 35-41

33. Lay C-Y, Mykytczuk NCS, Niederberger TD, Martineau C, Greer CW, Whyte LG (2012) Microbial diversity and activity in hypersaline high Arctic spring channels. Extremophiles 16: 177-191

34. Li X, Dobretsov S, Xu Y, Xiao X, Hung OS, Qian PY (2006) Antifouling diketopiperazines produced by a deep-sea bacterium, *Streptomyces fungicidicus*. Biofouling 22: 201-208

35. Li DL, Li XM, Proksch P, Wang BG (2010) 7-O-Methylvariecolortide A, a new spirocyclic diketopiperazine alkaloid from a marine mangrove derived endophytic fungus, *Eurotium rubrum*. Nat Prod Commun 5: 580-584

36. Llamas I, Quesada E, Canovas MJ, Gronquist M, Eberhard A, Gonzalez JE (2005). Quorum sensing in halophilic bacteria: detection of N-acyl-homoserine lactones in the exopolysaccharide-producing species of *Halomonas*. Extremophiles 9: 333-341

37. Lorenzen CJ (1966) A method for the continuous measurement of in vivo chlorophyll concentration. Deep Sea Res Oceanogr Abstr 13: 223-227

38. MacFaddin JF (2000) Biochemical tests for identification of medical bacteria. 3rd edn. Philadelphia, Lippincott Williams and Wilkins

39. Margesin R, Schinner F (2001) Potential of halotolerant and halophilic microorganisms for biotechnology. Extremophiles 5: 73-83

40. Montalvo-Rodríguez R, López-Garriga J, Vreeland RH, Oren A, Ventosa A, Kamekura M (2000) *Haloterrigena thermotolerans* sp. nov., a halophilic archaeon from Puerto Rico. Int J Syst Evol Microbiol 50: 1065–1071

41. Nithya C, Begum MF, Pandian SK (2010) Marine bacterial isolates inhibit biofilm formation and disrupt mature biofilms of *Pseudomonas aeruginosa* PAO1. Appl Microbiol Biotechnol 88: 341-358

42. Ojika M, Inukai Y, Kito Y, Hirata M, Iizuka T, Fudou R (2008) Miuraenamides: antimicrobial cyclic depsipeptides isolated from a rare and slightly halophilic myxobacterium. Chem Asian J 3: 126-133

43. Ortega-Morales BO, Chan-Bacab MJ, Miranda-Tello E, Fardeau M-L, Carrero JC, Stein T (2008) Antifouling activity of sessile bacilli derived from marine surfaces. J Ind Microb Biotech 35: 9-15

44. Ortlepp S, Pedpradap S, Dobretsov S, Proksch P (2008) Antifouling activity of sponge derived polybrominated diphenyl ethers and synthetic analogues. Biofouling 24: 201-208

45. Paggi RA, Martone CB, Fuqua C, De Castro RE (2003) Detection of quorum sensing signals in the haloalkaliphilic archaeon *Natronococcus occultus*. FEMS Microbiol Lett 221: 49-52

46. Park YC, Gunasekera SP, Lopez JV, McCarthy PJ, Wright AE (2006) Metabolites from the marine-derived fungus *Chromocleista* sp. isolated from a deep-water sediment sample collected in the Gulf of Mexico. J Nat Prod 69: 580–584

47. Pruesse E, Quast C, Knittel K, Fuchs BM, Ludwig W, Peplies J, Glöckner FO (2007) SILVA: a comprehensive online resource for quality checked and aligned ribosomal RNA sequence data compatible with ARB. Nucleic Acids Res 35: 7188–7196

48. Rasmussen TB, Givskov M (2006) Quorum-sensing inhibitors as anti-pathogenic drugs. Internat J Med Microb 296: 149-161

49. Rodriguez-Valera F, Ruiz-Berraquero F, Ramos-Cormenzana A (1980) Isolation of extremely halophilic bacteria able to grow in defined inorganic media with single carbon sources. J Gen Microbiol 119: 535-538

50. Sepcic K, Zalar P, Gunde-Cimerman N (2010) Low water activity induces the production of bioactive metabolites in halophilic and halotolerant fungi. Mar Drugs 9: 43-58

51. Qi S-H, Xu Y, Gao J, Qian P-Y, Zhang S (2009). Antibacterial and antilarval compounds from marine bacterium *Pseudomonas rhizosphaerae*. Ann Microb 59: 229-233

52. Sinha S, Srivastava R, De Clercq E, Singh RK (2004) Synthesis and antiviral properties of arabino and ribonucleosides of 1,3-dideazaadenine, 4-nitro-1,3-dideazaadenine and diketopiperazine. Nucleosides Nucleotides Nucleic Acids 23: 1815–1824

53. Sonnenschein EC, Gärdes A, Seebah S, Torres-Monroy I, Grossart H-P, Ullrich MS (2011) Development of a genetic system for *Marinobacter adhaerens* HP15 involved in marine aggregate formation by interacting with diatom cells. J Microbiol Methods 87: 176-183

54. Steppe TF, Olson JB, Paerl HW, Litaker RW, Belnap J (1996) Consortial $N_2$ fixation: a strategy for meeting nitrogen requirements of marine and terrestrial cyanobacterial mats. FEMS Microb Ecol 21: 149-156

55. Tahrioui A, Quesada E, Llamas I (2011) The hanR/hanI quorum-sensing system of *Halomonas anticariensis*, a moderately halophilic bacterium. Microbiology 157: 3378-3387

56. Tommonaro G, Abbamondi R, Iodice C, Tait K, De Rosa S (2012) Diketopiperazines produced by the halophilic Archaeon, *Haloterrigena hispanica*, activate AHL bioreporters. Microb Ecol 63: 490-495

57. Vahed SZ, Forouhandeh H, Hassanzadeh S, Klenk H-P, Hejazi MA, Hejazi MS (2011) Isolation and characterization of halophilic bacteria from Urmia Lake in Iran. Microbiology 80: 834-841

58. Van der Merwe E, Hunag D, Peterson D, Kilian G, Milne PJ, Van de Venter M, Frost, C (2008) The synthesis and anticancer activity of selected diketopiperazines. Peptides 29: 1305-1311

59. Wang G, Dai D, Chen M, Wu H, Xie L, Luo X, Li X (2010) Two diketopiperazine cyclo(pro-phe) isomers from marine bacteria *Bacillus subtilis* sp. 13-2. Chem Nat Comp 46: 583-585

60. Watts KR, Ratnam J, Ang K-H, Tenney K, Compton EJ, McKerrow J, Crews P (2010) Assessing the trypanocidal potential of natural and semi-synthetic diketopiperazines from two deep water marine-derived fungi. Bioorg Med Chem 18: 2566-2574

61. Winson MK, Swift S, Fish L, Throup JP, Jorgensen F, Chhabra SR, Bycroft BW, Williams P, Stewart GS (1998) Construction and analysis of luxCDABE-based plasmid sensors for investigating N-acyl homoserine lactone-mediated quorum sensing. FEMS Microbiol Lett 163: 185-192

62. Xu X-W, Liu S-J, Tohty D, Oren A, Wu M, Zhou P-J (2005) *Haloterrigena saccharevitans* sp. nov., an extremely halophilic archaeon from Xin-Jiang, China. Int J Syst Evol Microbiol 55: 2539-2542

63. Yebra DM, Kiil S, Weinell CE, Dam-Johansen K (2006) Effects of marine microbial biofilms on the biocide release rate from antifouling paints - A model-based analysis. Progr Org Coat 57: 56-66

64. Yonezawa K, Yamada K, Kouno I (2011) New diketopiperazine derivates isolated from sea urchin derived *Bacillus* sp. Chem Pharm Bull (Tokyo) 59: 106-108

65. Zar JH (1996) Biostatistical analysis. 3 rd edn. Upper Saddle River, Prentice Hall International.

**Figure legends**

**Table 1.** Morphological and biochemical properties of halophilic isolates

**Table 2**. Anti-bacterial activity of extracts (hexanes, dichloromethane, butanol, and water) and supernatants from the studied isolates against pathogens *Bacillus subtilis*, *Escherichia coli*, *Micrococcus luteus*, *Proteus vulgaris*, *Pseudomonas aeruginosa*, *Salmonella enterica*, *Shigella sonnei*, *Staphylococcus aureus* and *Streptococcus pyogenes*. Inhibition of growth of these pathogens was investigated by the disk diffusion bioassay using extracts and supernatants. Inhibition values are presented as mean (mm) ± standard deviation (n=3). All extracts were tested at 3 μg. Controls (hexanes, dichloromethane, butanol, and distilled water) did not affect the growth of pathogens (data are not shown). Bioassays were repeated in triplicate. The data significantly different ($p<0.05$) from the corresponding controls according to the Dunnet test have asterisks.

**Table 3**. Inhibition of QS activity of *Chromobacterium violaceum* CV017 by extracts and supernatants of halophilic strains. Data are expressed as the mean minimum inhibitory concentration (MIC, μg/ ml) ± standard deviation. Bioassays were repeated in triplicate. Control (DMSO) did not affect QS production of the reporter strain (data are not shown). The data significantly different ($p<0.05$) from the corresponding controls according to the Dunnet test have asterisks. The most active extracts are highlighted.

**Figure 1.** 16S rRNA-based phylogenetic reconstruction of the six halophilic strains isolated from a hypersaline microbial mat. An archaean part (grey) of the tree separated from a bacterial part (white) by background colour.

**Figure 2**. The effect of extracts and supernatants of halophilic strains on growth of the diatom *Amphora coffeaeformis* (A and B) and the green alga *Dunaliella salina* (C and D). The data presented as means + standard deviation of chlorophyll *a* concentrations in the presence or absence (control) of extracts. Solvents were hexanes, dichloromethane (DCM), butanol (BuOH), and distilled water ($H_2O$). Extracts from supernatants (A and C) and cell pellets (B and D) were tested. The data significantly different ($p<0.05$) from the corresponding controls according to the Dunnet test have asterisks.

**Figure 3**. Chemical structures of diketopiperazines isolated from SK-3. (**1**) *cyclo*(L-Pro-L-Phe), (**2**) *cyclo*(L-Pro-L-Leu), (**3**) *cyclo*(L-Pro-L-*iso*Leu) and (**4**) *cyclo*(L-Pro-D-Phe)

**Figure 4**. The effect of diketopiperazines isolated from *Marinobacter* sp. SK-3 on QS dependent bioluminescence of the LuxR-based reporter *E.coli* pSB401 induced by 3-oxo-C6-HSL (final concentration of 10 μM). (**A**) *cyclo*(L-Pro-L-Phe), (**B**) *cyclo*(L-Pro-L-*iso*Leu), (**C**) *cyclo*(L-Pro-L-Leu), and (**D**) *cyclo*(L-Pro-D-Phe). Data are means + SD relative bioluminescence (bioluminescence/$OD_{595}$) of the reporter with added compounds (n = 8). Compound concentrations that significantly (Dunnet, $p<0.05$) inhibited QS of the reporter are marked with asterisks. Positive control (PC) contained AHL and negative control (NC) did not contain AHL. All treatments and controls contained DMSO. Measurements were taken every 1 h but results obtained at 4h are shown since the inhibitory trend for the tested compounds did not change over time.

**Figure 1**

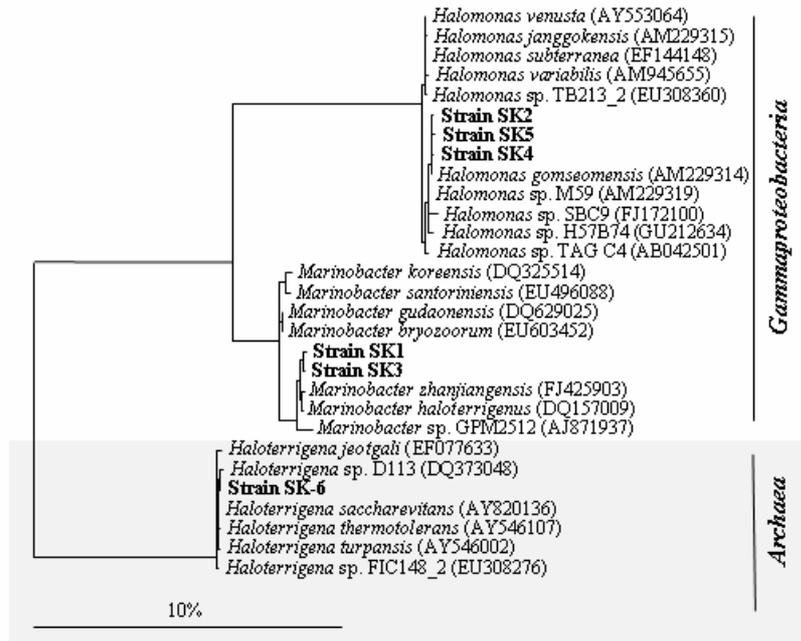

**Table 1**

| Characteristic | Isolates | | | | | |
|---|---|---|---|---|---|---|
| | SK-1 | SK-2 | SK-3 | SK-4 | SK-5 | SK-6 |
| *Morphology* | rod Pleomorphic | rod Pleomorphic | rod Pleomorphic | rod Pleomorphic | rod Pleomorphic | rod Pleomorphic |
| Gram stain | - | - | - | - | - | - |
| Pigmentation | white | yellow | white | yellow | yellow-brown | pink |
| Flagella | Yes | Yes | Yes | Yes | Yes | No |
| *Physiology* | | | | | | |
| Salinity (%) | 5-22 | 5-22 | 5-22 | 5-25 | 5-25 | 15-25 |
| Optimum salinity | 15 | 15 | 15 | 10 | 10 | 22 |
| Temperature (℃) | 15-45 | 15-45 | 15-45 | 15-45 | 15-45 | 20-60 |
| Optimum temperature | 35 | 35 | 35 | 35 | 25 | 45 |
| Catalase | + | - | + | - | - | - |
| Acid from: | | | | | | |
| D-Glucose | - | - | - | + | + | + |
| Lactose | - | + | - | + | - | + |
| Maltose | - | - | - | + | - | + |
| Sucrose | - | - | - | - | - | - |
| Hydrolysis of: | | | | | | |
| Starch | - | - | - | - | - | - |
| Gelatin | - | - | - | - | - | - |
| Tween 20 | + | + | + | + | + | - |
| CMC | + | + | + | - | - | - |
| Birchwood xylan | - | + | - | + | + | - |
| Indole | - | - | - | - | - | - |
| MR | - | - | - | - | - | - |
| VP | - | - | - | - | - | - |
| Urea | - | - | - | - | - | - |
| Citrate | - | - | - | - | - | - |
| Gelatin liquefaction | - | - | - | - | - | - |
| TSI | - | - | - | - | - | - |
| *Enzymes* | | | | | | |
| Lipase | + | + | + | + | + | - |
| Cellulase | + | + | + | - | - | - |
| Xylanase | - | + | - | + | + | - |
| Amylase | - | - | - | - | - | - |
| Protease | - | - | - | - | - | - |

CMC: Carboxymethylcellulose
MR: Methyl red
VP: Voges-Proskauer
TSI: Triple Sugar Iron

**Table 2**

| Strain | Pathogens | Hexanes | | Dichloromethane | | Butanol | | Water | |
|---|---|---|---|---|---|---|---|---|---|
| | | Supernatant | Pellet | Supernatant | Pellet | Supernatant | Pellet | Supernatant | Pellet |
| SK 1 | S. aureus | - | - | - | 1.0±0.1* | - | - | - | - |
| | S. pyogenes | - | - | - | - | - | 1.0±0.3* | - | - |
| | E. coli | - | - | - | - | - | - | - | - |
| | S. enterica | - | - | - | - | - | - | - | - |
| | S. sonnei | - | - | - | - | - | - | - | - |
| | M. luteus | - | - | - | - | - | - | - | - |
| | B.subtilis | 1.0±0.5* | 0.5±0.5 | - | - | - | - | - | - |
| | P.aeruginosa | - | - | - | - | - | - | - | - |
| | P. vulgaris | - | - | - | - | - | - | - | - |
| SK 2 | S. aureus | - | - | 4.0±1.0* | 6.0±1.1* | - | - | - | - |
| | S. pyogenes | - | - | - | - | - | 1.0±0.3* | - | - |
| | E. coli | - | - | - | - | - | - | - | - |
| | S. enterica | - | - | - | - | - | - | - | - |
| | S. sonnei | - | - | - | - | - | - | - | - |
| | M. luteus | - | - | - | - | - | - | - | - |
| | B.subtilis | 2.5±0.5* | 0.5±0.5 | - | - | - | - | - | - |
| | P.aeruginosa | - | - | - | - | - | - | - | - |
| | P. vulgaris | - | - | - | - | - | - | - | - |
| SK 3 | S. aureus | - | - | 3.0±0.0* | 3.0±1.0* | - | - | - | - |
| | S. pyogenes | - | - | - | - | - | 2.0±1.0* | - | - |
| | E. coli | - | - | - | - | - | - | - | - |
| | S. enterica | - | - | - | - | - | 1.5±0.5* | - | - |
| | S. sonnei | - | - | - | - | - | - | - | - |
| | M. luteus | - | - | - | - | - | - | - | - |
| | B.subtilis | 0.5±0.5 | 0.9±0.3* | - | - | - | 2.5±0.9* | - | - |
| | P.aeruginosa | - | - | - | - | - | - | - | - |
| | P. vulgaris | - | - | - | - | - | - | - | - |
| SK 4 | S. aureus | - | - | 3.0±1.0* | 2.0±0.3* | - | - | - | - |
| | S. pyogenes | - | - | - | - | - | 2.0±1.0* | - | - |
| | E. coli | - | - | - | - | - | - | - | - |
| | S. enterica | - | - | - | - | - | - | - | - |
| | S. sonnei | - | - | - | - | - | - | - | - |
| | M. luteus | - | - | - | - | - | - | - | - |
| | B.subtilis | 0.7±0.7 | 0.5±0* | - | - | - | - | - | - |
| | P.aeruginosa | - | - | - | - | - | - | - | - |
| | P. vulgaris | - | - | - | - | - | - | - | - |
| SK 5 | S. aureus | - | - | 0.5±0.5 | 4.0±1.0* | - | - | - | - |
| | S. pyogenes | - | - | - | - | - | 2.0±1.0* | - | - |
| | E. coli | - | - | - | - | - | - | - | - |
| | S. enterica | - | - | - | - | - | - | - | - |
| | S. sonnei | - | - | - | - | - | - | - | - |
| | M. luteus | - | - | - | - | - | - | - | - |
| | B.subtilis | 4.0±1.0* | 0.5±0.5 | - | - | - | - | - | - |
| | P.aeruginosa | - | - | - | - | - | - | - | - |
| | P. vulgaris | - | - | - | - | - | - | - | - |
| SK 6 | S. aureus | - | - | 4.0±1.0* | 3.3±1.5* | - | - | - | - |
| | S. pyogenes | - | - | - | - | - | 2.0±0.0* | - | - |
| | E. coli | - | - | - | - | - | - | - | - |
| | S. enterica | - | - | - | - | - | - | - | - |
| | S. sonnei | - | - | - | - | - | - | - | - |
| | M. luteus | - | - | - | - | - | - | - | - |
| | B.subtilis | 3.0±1.0* | 2.0±0.0* | - | - | - | - | - | - |
| | P.aeruginosa | - | - | - | - | - | - | - | - |
| | P. vulgaris | - | - | - | - | - | - | - | - |

- Absence of anti-bacterial activity

**Figure 2**

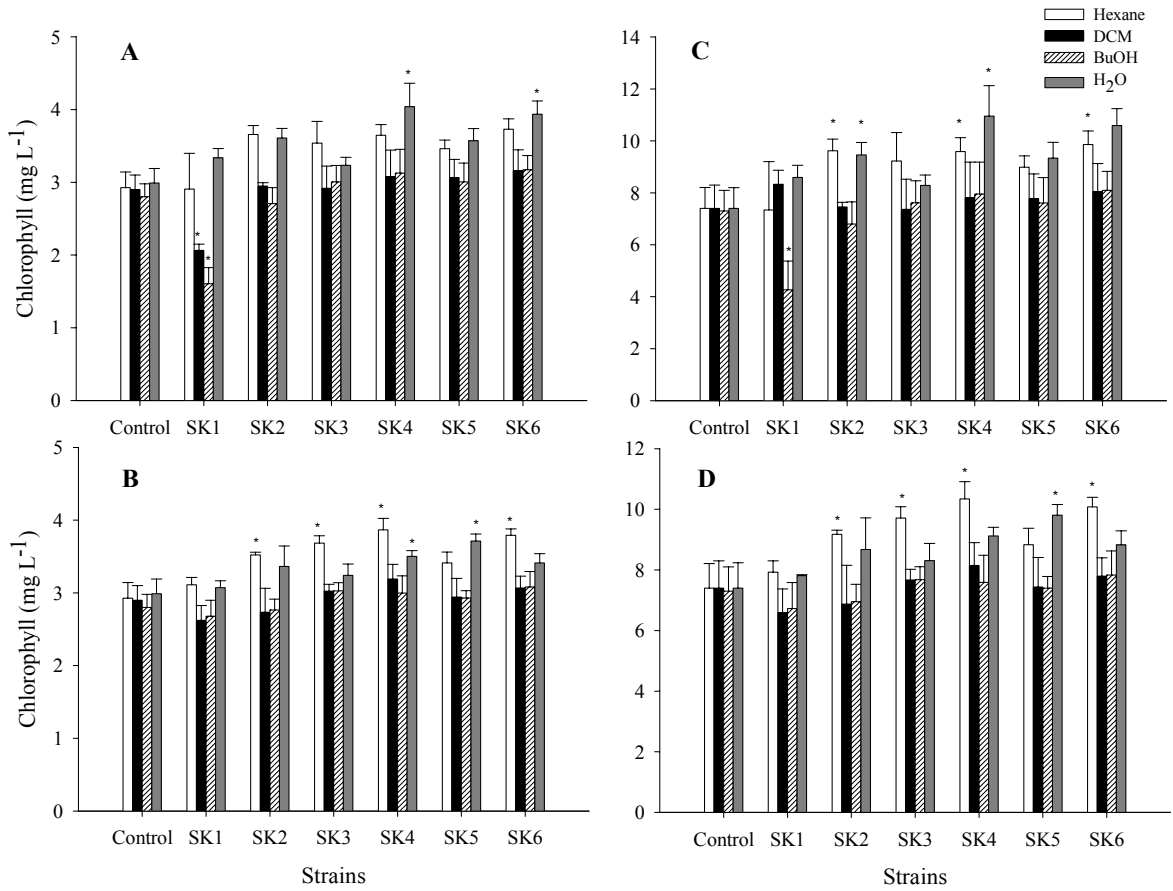

**Table 3**

| Strain | Hexane | | Dichloromethane | | Butanol | | Water | |
|---|---|---|---|---|---|---|---|---|
| | Supernatant | Pellet | Supernatant | Pellet | Supernatant | Pellet | Supernatant | Pellet |
| SK-1 | 1.10 ±0.10* | 10.60 ±1.50* | 10.00±1.00* | 9.60±1.50* | 10.30±0.50* | 10 ±1.00* | 0.37 ±0.04* | 10.00±1.00* |
| SK-2 | - | 9.60 ±0.50* | 0.37±0.04* | 10.30±0.60* | 0.33±0.04* | - | 1.10±0.10* | 3.30± 0.30* |
| SK-3 | 10.00±0.00* | 3.30 ±0.60* | 0.37±0.01* | 0.37 ±0.00* | 0.12±0.01* | 0.04 ±0.01* | 0.37 ±0.03* | 3.30± 0.30* |
| SK-4 | 3.30±0.60* | 3.30 ±0.60* | 0.37±0.02* | - | 0.12 ±0.00* | 0.12 ±0.01* | 0.37 ±0.01* | 10.00± 1.00* |
| SK-5 | 10.00±1.00* | - | 0.12±0.01* | 0.04±0.01* | 0.12 ±0.01* | 0.12 ±0.01* | 0.03 ±0.01* | 0.12± 0.01* |
| SK-6 | 3.30±0.60* | 10.30±1.50* | 1.10 ±0.30* | - | 10.30±0.60* | 10 ±0.00* | 1.1 ±0.10* | 10.30±1.50* |

- Absence of quorum sensing inhibition

Figure 3

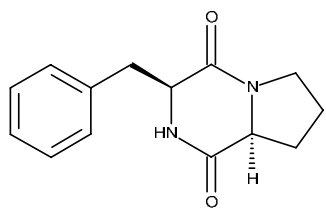
(1)

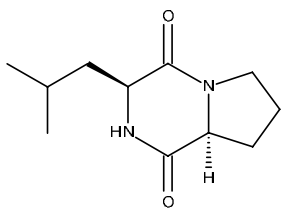
(2)

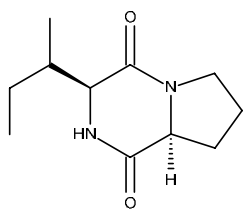
(3)

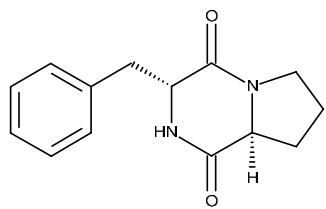
(4)

**Figure 4**

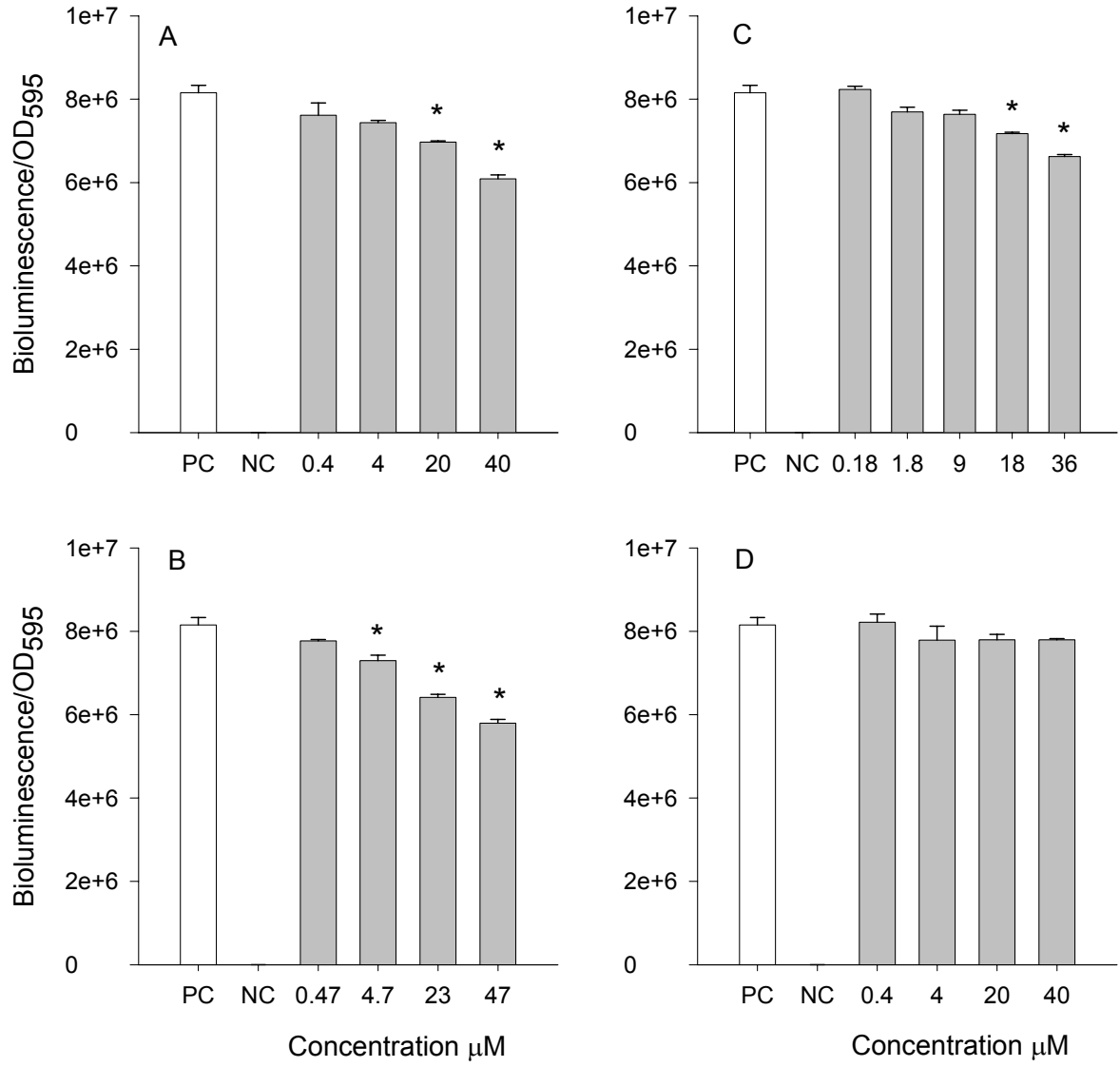